\documentclass[preprint]{elsarticle}
\usepackage{amssymb}
\usepackage{footnote} %para footnote en tabla (entorno savenotes)

\begin{document}
\begin{frontmatter}
\title{Study of scintillation in natural and synthetic quartz and methacrylate}
\author{J.~Amar\'e,
%\author[uz,lsc]{
S.~Borjabad, S.~Cebri\'an\footnote{Corresponding author},
C.~Cuesta\footnote{Present address: Center for Experimental Physics
and Astrophysics, University of Washington, Seattle, WA, US},
D.~Fortu\~{n}o\footnote{Present address: ENDESA Generaci\'on,
Lleida, Spain}, E.~Garc\'ia, C.~Ginestra,
H.~G\'omez\footnote{Present address: Laboratoire de
l'Acc\'el\'erateur Lin\'eaire (LAL). Centre Scientifique d'Orsay.
B\^{a}timent 200 - BP 34. 91898 Orsay Cedex, France.}, D.C.~Herrera,
M.~Mart\'inez, M.A.~Oliv\'an, Y.~Ortigoza, A.~Ortiz de Sol\'orzano,
C.~Pobes\footnote{Present address: Instituto de Ciencia de
Materiales de Arag\'on, Universidad de Zaragoza - CSIC},
J.~Puimed\'on, M.L.~Sarsa, J.A.~Villar and P.~Villar}
\address{Laboratorio de F\'isica Nuclear y Astropart\'iculas, Universidad de Zaragoza\\
Calle Pedro Cerbuna 12, 50009 Zaragoza, Spain \\
%\address[lsc]{
Laboratorio Subterr\'aneo de Canfranc\\
Paseo de los Ayerbe s/n, 22880 Canfranc Estaci\'on, Huesca, Spain}
%\date{}
%\maketitle
%\fntext[carlos]{Present address:}

\begin{abstract}
Samples from different materials typically used as optical windows
or light guides in scintillation detectors were studied in a very
low background environment, at the Canfranc Underground Laboratory,
searching for scintillation. A positive result can be confirmed for
natural quartz: two distinct scintillation components have been
identified, not being excited by an external gamma source. Although
similar effect has not been observed neither for synthetic quartz
nor for methacrylate, a fast light emission excited by intense gamma
flux is evidenced for all the samples in our measurements. These
results could affect the use of these materials in low energy
applications of scintillation detectors requiring low radioactive
background conditions, as they entail a source of background.
%poner taus?
\end{abstract}

\end{frontmatter}

\section{Introduction}

Scintillation detectors are used, among many other applications, in
rare event searches as the study of the neutrinoless double beta
decay \cite{elliott} or the direct detection of hypothetical dark
matter particles pervading our galactic halo \cite{baudis}.
Improving the background levels and energy thresholds of
scintillation detectors is still a challenging issue.
Photomultiplier Tubes (PMTs) were typically a strong background
source, however, ultra-low background models are becoming state of
the art, and then, other contributions to the background become more
relevant \cite{PMTdama,PMTxe}. In the case of inorganic
scintillators, being NaI(Tl) one with the most outstanding
performance, PMTs are typically coupled to the scintillator crystal
through light guides or optical windows made of different materials,
whose design is based on geometry or background considerations. In
the development of prototypes for the ANAIS (Annual modulation with
NaI Scintillators) experiment \cite{anais} operating in the Canfranc
Underground Laboratory (LSC, Laboratorio Subterr\'aneo de Canfranc,
Spain) and using NaI(Tl) crystals as target material, understanding
of the different background contributions has been a major issue:
residual radioactive backgrounds have been studied in
\cite{anaisbkg}, a very slow scintillation time constant in NaI(Tl)
able to trigger the experiment and then, contribute to the
background, was identified recently \cite{slowscint}, and the
results of the scintillation study of different materials considered
to be used as optical windows and light guides are presented in this
letter. This study was motivated by the identification, in previous
ANAIS prototype tests using natural quartz optical windows, of a
population of events, which contributed to the background at very
low energy, and whose pulses showed an abnormal temporal behavior,
since their decay was faster than expected for NaI(Tl) scintillation
events. This letter describes a specific mounting and the set of
measurements carried out in order to confirm and characterize a
possible scintillation in natural quartz, and to test, under similar
conditions, other suitable materials for the same purpose, namely
synthetic quartz and methacrylate, to be sure that a similar effect
is not present and they could be used in dark matter experiments.
This study is not directly related with the dark matter search goal
of ANAIS experiment, in whose context natural quartz was directly
discarded as optical window because of the high radioactive content
of the material. However, scintillation in natural quartz had not
been previously related with fast populations of events in NaI(Tl)
detectors, and in our opinion, this issue deserved more
understanding.

%descripcion general PMT noise
Events produced by a weak scintillation in the materials searched
for would appear entangled with dark events from PMTs, being
difficult to discriminate. This PMT noise is mainly due to
thermoionic electron emission, but other contributions can be
present due, for instance, to the emission of Cherenkov light
(generated by radioactive contamination in the PMT, in their
neighborhood or, even, by the interaction of cosmic rays) and ions
in residual gas inside the PMT \cite{knoll}. Operating two PMTs in
coincidence is a common and effective practice in experiments
demanding a low energy threshold in order to diminish the
contribution from the thermoionic effect. However, Cherenkov light
emission can produce coincident events, although they are expected
to be quite asymmetric. Residual gas ions typically produce
afterpulses, not relevant for coincidence measurements. The residual
noise when operating PMTs in coincidence, as it is done in this
work, attributed to accidental coincidences from dark events and to
the generation of light in one PMT being detected by the other, had
been studied in \cite{robinson}.

Quartz is not considered as a scintillating material in the
literature \cite{librobirks,lecoq}; a weak scintillation was
reported under irradiation with $\alpha$ particles
\cite{librobirks,birks}, although it was not characterized. However,
quartz is largely used for dosimetry and dating thanks to thermal
and optically stimulated luminescence, since it can emit when heated
or illuminated an amount of light proportional to the radiation dose
accumulated in time \cite{aitken}. Imperfections (impurities or
defects) in crystalline materials (as quartz) disturb the
periodicity of the crystalline electric field, generating dips in
the electric potential where free electrons may be trapped (the
so-called electron traps). Ionizing radiation is able to excite
electrons into the conduction band, that eventually could fall in
such traps. Hence, the amount of electrons trapped is proportional
to the radiation dose received. In the case of materials used for
dosimetry or dating, when the irradiated material is heated or
exposed to strong light, trapped electrons can absorb enough energy
to return into the conduction band and then, recombine with holes in
the valence band emitting part of the energy in the form of
luminescence. In the case of quartz, used for optical dating, green
or blue light is used to free the trapped electrons and luminescence
in the ultraviolet is produced and read out. Then, scintillation in
the ultraviolet range is expected from quartz also directly
following energy deposition from ionizing radiation, in a time scale
dependent on the lifetime of the radiative decaying states
participating in the scintillation mechanism and with, probably,
very low intensity. Characterization and understanding of such a
scintillation was the main goal of this work.

In section \ref{mea} the experimental setup and the plan of
measurements carried out to study possible scintillation in
different quartz and methacrylate samples, to be used as optical
windows and/or light guides with NaI(Tl) crystals, are described.
Analysis performed and results obtained are presented in section
\ref{res}, considering first the measurements made without sample
and then those with the different samples. Section \ref{conclusions}
gathers the conclusions.
% and finally, going in depth into the natural quartz case.

\section{Experimental set-up and summary of measurements}
\label{mea}

\begin{figure}
 \begin{center}
 \includegraphics[width=10cm]{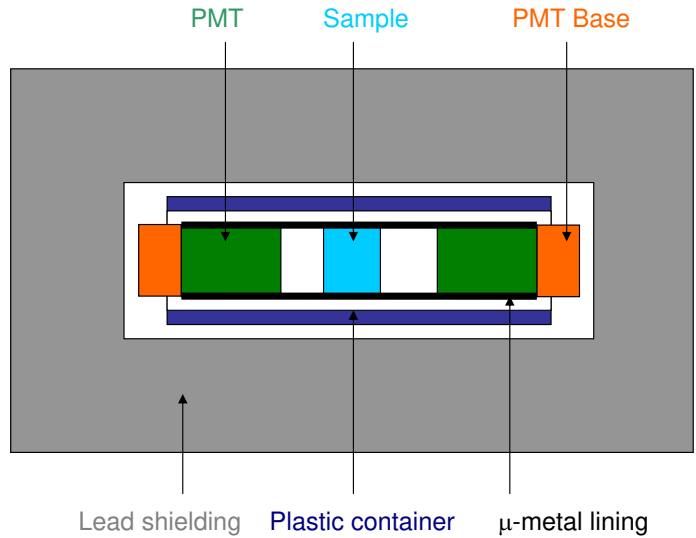}
 \end{center}
 \caption{Sketch of the scintillation test bench used for the measurements presented in this work: a plastic cylinder houses the scintillator sample and two PMTs lined in $\mu$-metal, inside a lead shielding at the Canfranc Underground Laboratory.}
  \label{photo}
\end{figure}

%descripici\'on del set-up
A test bench was conditioned at LSC to study scintillation from
different samples. It consisted of two photomultipliers (Electron
Tubes 9302B model, 3~inches diameter) faced at a fixed 10~cm
distance between their respective photocathodes, in order to keep as
much as possible the same geometry in the different measurements.
The samples were placed centered in the inner space, between the
PMTs (see figure~\ref{photo}), without using optical grease for the
coupling to the PMTs to minimize the incorporation of unnecessary
components that could affect the comparison between samples.
Photomultipliers were operated at 1100~V. According to the 9302B
series data sheet, the transit time is 40~ns and the single electron
response (SER) rise time is 7.5~ns and 15~ns the corresponding FWHM.
The nominal dark count rate at 20$^{o}$C is 500~s$^{-1}$. The
spectral response curve ranges from 300 to 500~nm, having the
maximum at 350-400~nm. Charge readout of the PMT is fast enough to
keep the single electron time behavior. Samples and PMTs were housed
in a 0.1-mm-thick $\mu$-metal lining, covered by a plastic container
made of PVC to avoid environmental light reaching the PMTs. Since
the dark counting rate of PMTs can be affected by environmental
gamma or cosmic ray fluxes, the set-up was operated in low
background conditions: a 10-cm-thick lead shielding was used to
reduce the contribution from environmental gamma radiation and all
measurements were carried out at LSC, located in the Spanish
Pyrenees, under a rock overburden of 2450~m.w.e. Underground
operation at such a depth guarantees a significant cosmic ray
suppression; the measured muon flux at LSC is of the order of
10$^{-7}$~cm$^{-2}$s$^{-1}$ \cite{luzon,bettini}, which means a
reduction of about five orders of magnitude with respect to the flux
above ground.

\begin{savenotes}
\begin{table}
\begin{center}
\caption{Main features of the samples studied in this work:
material, supplier, dimensions and measured activities (or upper
limits) of $^{232}$Th and $^{238}$U natural chains using HPGe
spectrometry at LSC \cite{anaisbkg}.} \centering
\begin{tabular}{|l|c|c|c|c|c|c|c|}
\hline
 Material &   Supplier & Diameter  & Height  & $^{232}$Th  & $^{238}$U & $^{40}$K & units \\
 & & (mm) & (mm) & &  &  & \\
  \hline
Natural quartz & Heraeus & 76.4 & 10 & 33$\pm$4  &
228$\pm$9 & $<$50 &mBq/kg\\
(Homosil) &&&& &  & &\\ \hline
Synthetic quartz & Heraeus & 76.2 & 10 & $<$4.7 $^{228}$Ra  & $<$100 $^{238}$U & $<$12 &mBq/kg\\
 (Suprasil 2 grade B)&&&& $<$2.5 $^{228}$Th & $<$1.9  $^{226}$Ra & &\\ \hline
Methacrylate  & Goodfellow & 78.0 & 100 & $<$4.5 $^{228}$Ra & $<$120 $^{238}$U & $<$21 &mBq/guide \\
(PMMA)\footnote{Screening of the methacrylate sample was carried out
including also a copper can.}& & & & $<$5.0 $^{228}$Th & $<$4.7  $^{226}$Ra & & \\
\hline
\end{tabular}
 \label{samples}
\end{center}
\end{table}
\end{savenotes}

Three different samples have been studied in this test bench:
natural and synthetic quartz and methacrylate. Synthetic and natural
quartz samples are from Suprasil and Homosil series from Heraeus,
respectively. According to supplier specifications, for Homosil
quartz only carefully selected crystal is used as raw material; it
has no particulate structure and extremely favorable homogeneity
properties. Suprasil 2 is a high purity synthetic fused silica
material manufactured by flame hydrolysis of SiCl$_{4}$; the index
homogeneity is controlled and specified in one direction. Other
relevant information about the samples is given in
table~\ref{samples}; for quartz samples, two identical cylinders as
those described in table \ref{samples} were placed together in the
bench. The samples were screened for radiopurity at the LSC using an
ultra-low background HPGe detector in order to determine the
corresponding activity of the $^{232}$Th and $^{238}$U natural
chains and $^{40}$K \cite{anaisbkg}. Results are shown in
table~\ref{samples}.
%this information will be useful for the interpretation of the obtained results.
For the natural quartz samples, data are compatible with equilibrium
in both chains; for synthetic quartz and methacrylate, upper limits
are given for $^{238}$U, $^{226}$Ra, and for the long-lived
daughters of $^{232}$Th.

%descripci\'on de electr\'onica
%No hay m\'as info en anais.cfg, TDS.cfg
%Ventana de coincidencias: PII (tesis Mar\'ia) 120 ns, tesis Carlos 100 ns
Data acquisition directly ran at a Tektronix oscilloscope (Digital
Phosphor Oscilloscope, TDS5034B) having 4 channels and a bandwidth
of 350~MHz. Pulse shapes were digitized in a window of 2~$\mu$s
taking 2500~samples. Trigger was done at photoelectron level in
logical AND between the two PMT signals in order to reduce dark
events. Time and the two digitized PMT pulses were recorded for each
event; we will refer below to both digitized PMT signals as 0 and 1,
respectively.

%medidas
Table~\ref{measurements} summarizes all the measurements performed,
including live time and trigger rates registered. For every sample,
two measurements were carried out, without external excitation (we
will refer to as background measurement) and under exposure to gamma
radiation produced by an external $^{232}$Th source with activity of
$\sim$5 kBq placed inside the shielding. Special measurements
without sample were also made, both in background conditions and
using the $^{232}$Th source, and even using a black cardboard
between the PMTs to suppress possible effects from light emission in
a PMT generating a coincident event in the other one.

\begin{table}
\begin{center}
\caption{Summary of measurements performed in the scintillation test
bench at LSC: sample type, exposure to an external gamma $^{232}$Th
source, live time and trigger rate.} \centering
\begin{tabular}{|c|c|c|c|}
\hline
 Sample &   External  & Live time  & Trigger rate  \\% & Rate after quality cuts \\
 & gamma source & (h)& (h$^{-1}$)\\ %& (h$^{-1}$)\\
  \hline
without sample & No & 79.60 & 921.4$\pm$3.4 \\ %& 632.9$\pm$2.8\\
 natural quartz & No & 319.11 & 1218.0$\pm$2.0\\ %&979.5$\pm$1.8\\
  synthetic quartz & No & 314.41 & 940.4$\pm$1.7\\ %& 744.1$\pm$1.5\\
 methacrylate & No & 908.33 & 621.38$\pm$0.83\\ %&468.1$\pm$0.7\\
 \hline
 without sample &  Yes & 0.52 & (12.79$\pm$0.16)$\times10^{3}$\\ %&(8.98$\pm$0.13)$\times10^{3}$\\
 natural quartz & Yes & 0.83 & (79.69$\pm$0.31)$\times10^{3}$\\ %&(69.04$\pm$0.28)$\times10^{3}$\\
 synthetic quartz & Yes & 2.08 & (70.74$\pm$0.19)$\times10^{3}$ \\ %&(57.55$\pm$0.17)$\times10^{3}$\\
 methacrylate & Yes & 2.30 & (149.78$\pm$0.26)$\times10^{3}$\\ %& (117.53$\pm$0.23)$\times10^{3}$\\
 without sample, black  & Yes & 2.15 & 656$\pm$17\\ %& 568$\pm$16\\
cardboard between PMTs & & & \\ %&\\
\hline
\end{tabular}
 \label{measurements}
\end{center}
\end{table}

\section{Analysis and Results}
\label{res}

%definicion de parametros usados en el analisis
In a first analysis, a set of parameters was calculated for every
registered event and saved in ROOT \cite{root} files for further
analysis. Among these parameters, we will use in the following:
\begin{itemize}
\item The area of the pulse.
\item The so-called P1 defined as:
\begin{equation}
P1=\frac{Area(100-600ns)}{Area(0-600 ns)}
\end{equation}
%\begin{equation}
%P2=\frac{Area(0-50ns)}{Area(0-100 ns)}
%\end{equation}
where the beginning of the pulse defines the zero time. This
parameter has been successfully used to disentangle PMT origin
events from real scintillation events at very low energy in NaI(Tl)
detectors \cite{anais,dama}, profiting from the quite long
scintillation main time constant ($\sim$230~ns) vs typical PMT
pulses FWHM\footnote{The FWHM of the SER for PMT models used for
individual photon counting are typically below 30~ns.}. P1
distribution for noise is centered at 0 while for NaI(Tl)
scintillation events peaks typically at $\sim$0.7.
\item The number n of peaks identified in the pulse. Profiting from the good sampling rate of the digitized data,
the discrete arrival of the photons to the PMT photocathode can be
distinguished in the pulses corresponding to low energy events. An
algorithm that finds and counts the number of peaks in the pulse has
been developed, it it based on the TSpectrum ROOT class and the
Search method \cite{morhac}. Peaks are considered gaussian with a
minimum height and width, selected specifically according to the SER
of the PMT used.
\item The possible correlation of the light shared between the two PMT
signals will be studied by evaluating the asymmetry parameter:
\begin{equation}
Asy=\frac{Area_{signal 0}-Area_{signal 1}}{Area_{signal
0}+Area_{signal 1}} \label{Asy}
\end{equation}
%the ratio between the difference and the sum of the two PMT signal areas.
The distribution of this parameter depends on PMT gains, which were
adjusted to give similar responses at a few per cent level.
\end{itemize}

These parameters have been calculated independently for each PMT
signal, but when needed, they can be calculated for the sum pulse
(in the case of the P1 parameter we will refer to this new parameter
as P1s). In the following, we will study the distribution of these
parameters and their correlations in the events registered for all
the experimental runs summarized in table~\ref{measurements}, after
removing electronic noise (thanks to abnormal values of the maximum
of digitized pulses) and applying some quality cuts (events showing
baselines values out of range or negative areas due to analysis
artifacts are also disregarded), as further described in
section~\ref{mws}.

%\subsection{PMT noise}
\subsection{Measurements without sample}
\label{withoutsamples}

%COMPLETAR ------------------
%Hay que dar el tamaño de la ventana de la coincidencia para hacer una estimaci\'on de la posible contribuci\'on de fortuitas de este tipo.
%---------------------------------

\begin{figure}
 \begin{center}
\includegraphics[width=8cm]{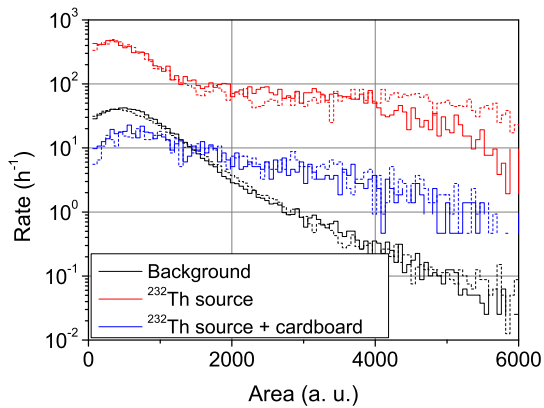}
 \end{center}
 \caption{Distributions (per hour) of the area of each PMT signal in all the measurements performed without sample (see table~\ref{measurements}): background measurement (black lines), measurement with
$^{232}$Th source (red lines) and with $^{232}$Th source and a black
cardboard between the PMTs (blue lines). Solid and dashed lines
correspond to each one of the two PMT signals.}
 \label{PMTsareas}
\end{figure}

\begin{figure}
 \begin{center}
   \includegraphics[width=8cm]{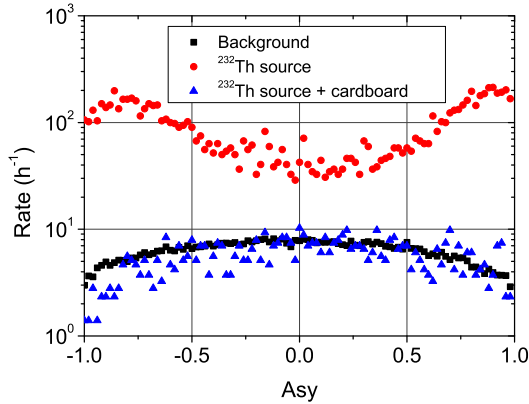}
 \end{center}
 \caption{Distributions (per hour) of the Asy parameter in all the measurements performed without sample: background, with $^{232}$Th source without black cardboard, and with
$^{232}$Th source and black cardboard.}
 \label{PMTsasymmetry}
\end{figure}

%nuestros resultados
The measurements performed in our scintillation test bench without
sample provided a population of PMT events. Figure~\ref{PMTsareas}
shows the distribution of the area of each PMT signal for the three
measurements performed: without the external gamma source, with the
source, and with the source and a black cardboard between the PMTs.
A possible correlation between the two PMT signals has been studied
by means of the Asy parameter, defined in eq.~(\ref{Asy});
distributions per hour of such a parameter are shown in
figure~\ref{PMTsasymmetry} for the three measurements without
sample.

%resultados con y sin fuente
The effect of gamma radiation on the PMTs is clearly evidenced by
the comparison of the rates in the background measurement and with
the $^{232}$Th source, both performed without sample. The trigger
rate increased under gamma excitation by a factor $\sim$14, as shown
in table~\ref{measurements}; this increase is more important for
events with larger areas (see figure~\ref{PMTsareas}). The gamma
flux from the $^{232}$Th source seems to generate a dominant
population of strongly asymmetric coincident events, as shown in
figure~\ref{PMTsasymmetry}.
%In the background measurement without sample this kind of asymmetric events are also present,
%although with much lower rate, together with the expected random
%coincidence events (see figure~\ref{PMTsareas}, top).
% in figure~\ref{noiseevent}, bottom the average pulse shapes of each PMT for the horizontal and vertical bands defined in the areas correlation plot are depicted separately.
% ; the average pulse shapes of each PMT shown in figure~\ref{noiseevent}, top left are quite similar, but asymmetric pulses have been identified too
%resultados con cartulina
To assess if this effect was related to the generation of light by
the gamma quanta, the measurement with a black cardboard between the
PMTs, preventing the light produced in one of them to reach the
other one, was carried out.  A clean sample of random coincidences
has been obtained in this case (see figure~\ref{PMTsasymmetry}). It
can be observed in table~\ref{measurements} how the trigger rate is
reduced a factor of 20 with respect to the measurement without
cardboard (being this rate even lower than the one without the
$^{232}$Th source), confirming the effect of generation of light in
a PMT being detected by the other one, described in \cite{robinson}.
Cherenkov light emission at PMT glass is possibly responsible for
these events. Following figure~\ref{PMTsasymmetry}, random
coincidences could explain a large fraction of the events in the
background measurement.
%For this measurement the average pulse shapes of
%each PMT shown in figure~\ref{noiseevent}, top right are similar and
%their temporal behavior agrees with the expectations.

\begin{figure}
 \begin{center}
    \includegraphics[width=8cm]{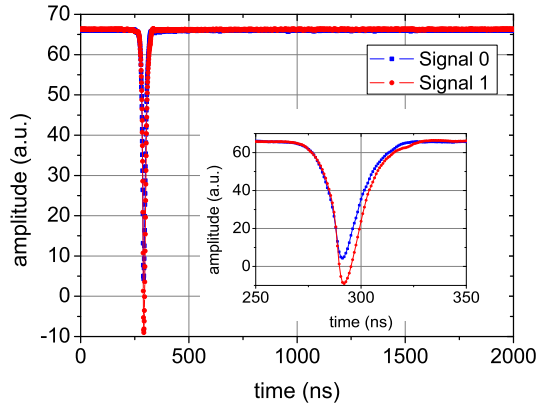}
 \end{center}
 \caption{Average pulse shapes of each PMT signal for random coincident events obtained in the measurement performed without sample, with  $^{232}$Th source and black cardboard. The whole digitized window as well as a zoom are displayed.}
  \label{noiseevent}
\end{figure}

%pulsos
Pulse shapes digitized in all the measurements without sample are
very similar. Figure~\ref{noiseevent} shows the average pulses of
each PMT for random coincident events obtained in the measurement
performed without sample, with $^{232}$Th source and black
cardboard. They show a gaussian shape with a 16 ns FWHM, in good
agreement with PMT specifications.

%% solo pulsos promedio?

\begin{figure}
 \begin{center}
 \includegraphics[width=8cm]{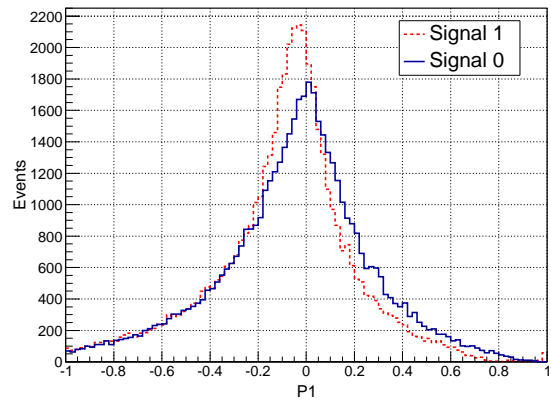}
 \includegraphics[width=8cm]{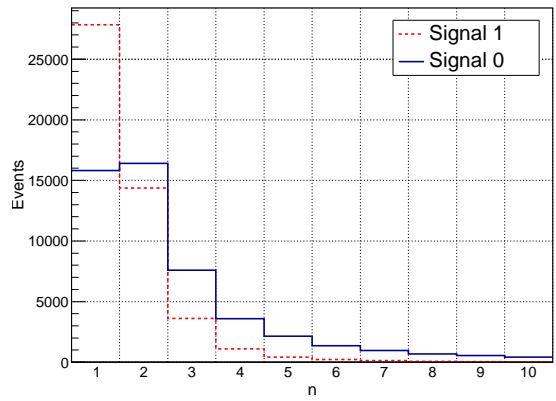}
 \end{center}
 \caption{Distributions of parameters P1 (top) and n (bottom) for the background measurement without sample. Distributions for signal 0 (blue solid lines) and 1 (red dashed lines) are independently shown.}
 \label{nP1}
\end{figure}

%%%%%%%%%%%%% distribuciones relevantes para filtrado de ruido
Figure~\ref{nP1} presents the distributions of the parameters P1 and
the number n of peaks identified per signal for the background
measurement without sample. P1 is distributed around 0 as expected
and the mean value of n is 2.71 and 1.54 for each one of the PMT
signals. These distributions will be helpful to fix cut values in
the event selection process implemented for the measurements with
samples, described in next section.

%\subsection{Identification of scintillation}
\subsection{Measurements with samples} \label{mws}

\begin{figure}
 \begin{center}
       \includegraphics[width=7cm]{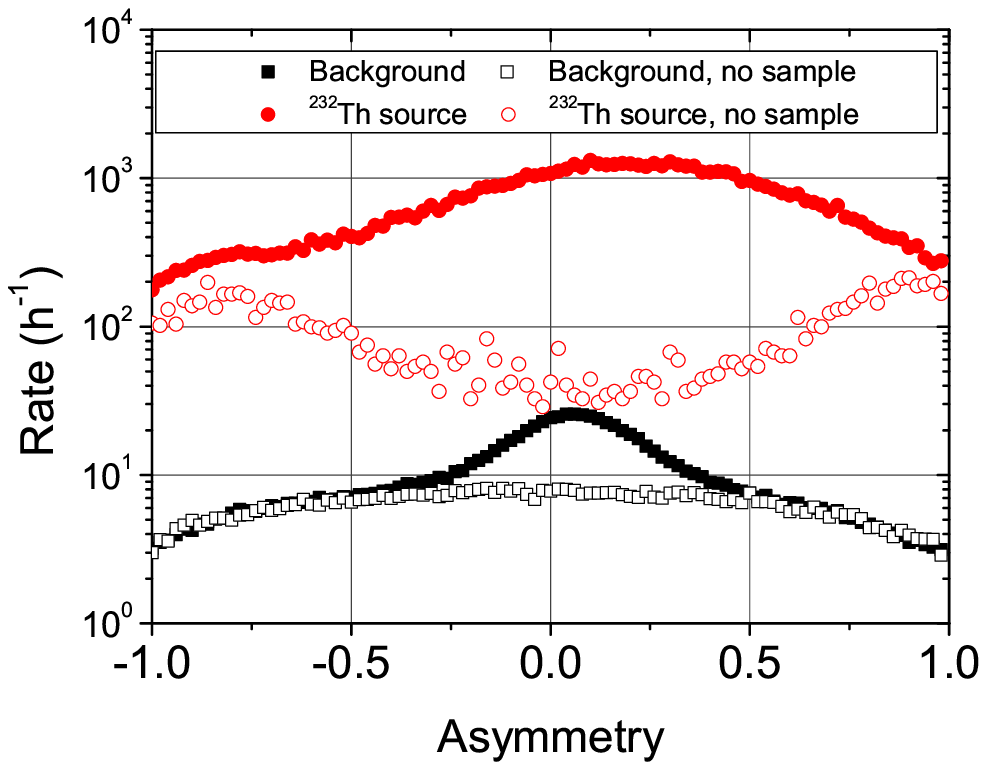}
    \includegraphics[width=7cm]{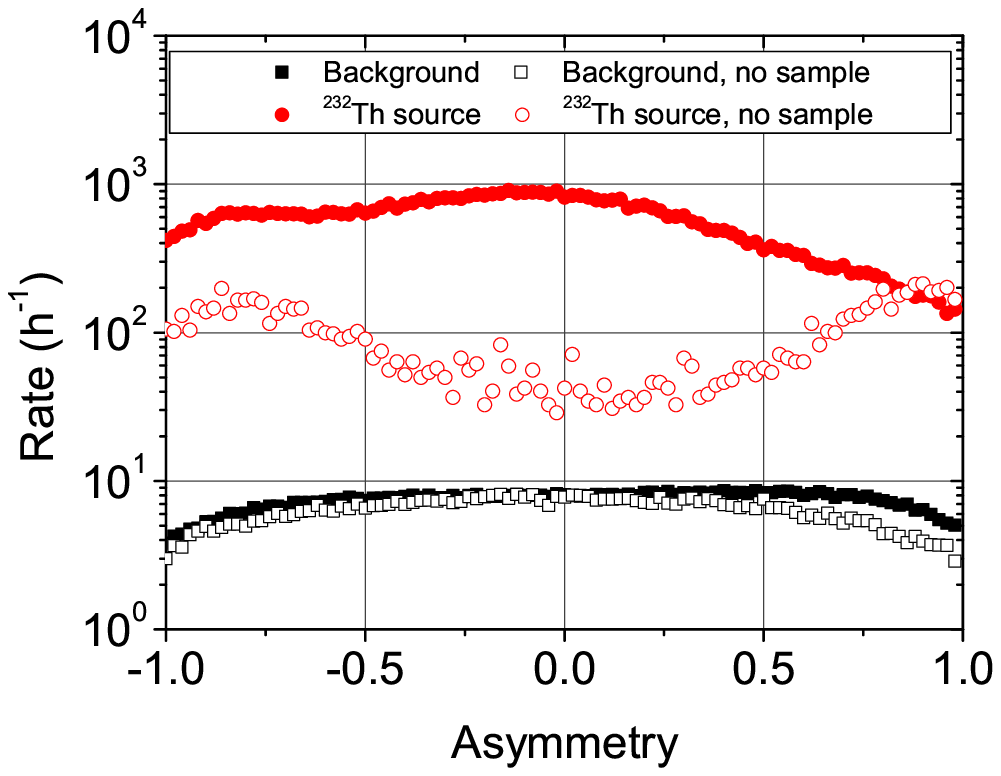}
  \includegraphics[width=7cm]{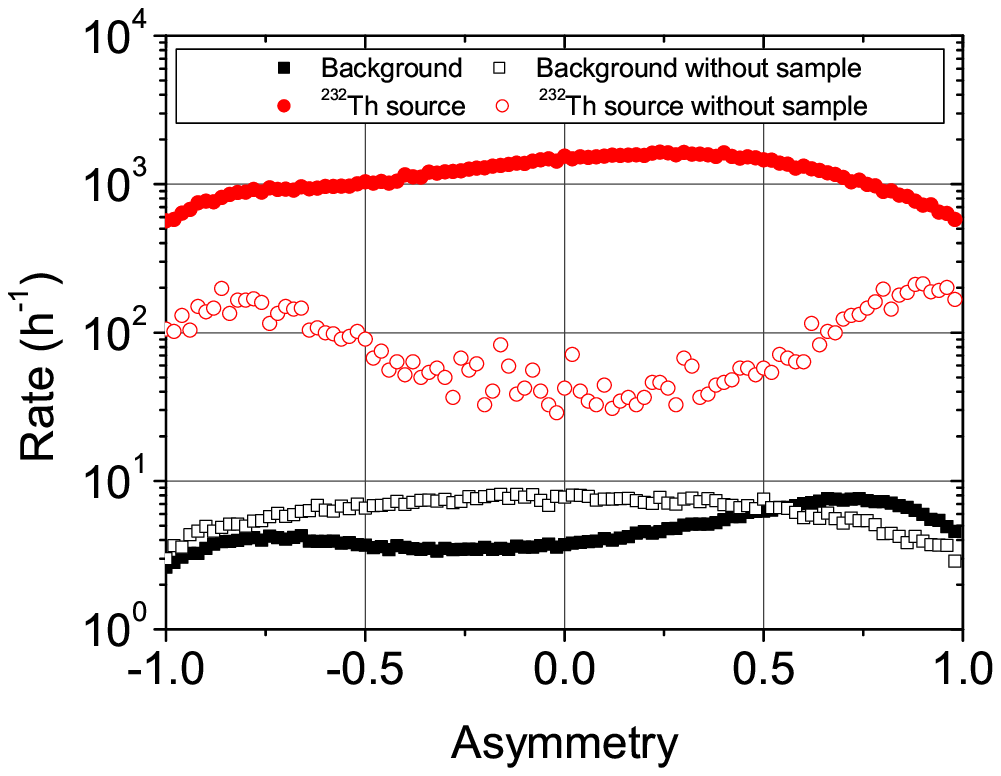}
 \end{center}
 \caption{Distributions (per hour) of the Asy parameter in the measurements with natural quartz, synthetic quartz and methacrylate (from top to bottom). Results for background (in black squares) and under exposure to a $^{232}$Th source (in red circles) are compared. Distributions corresponding to the measurements without sample shown in figure~\ref{PMTsasymmetry} are included also here for comparison (empty markers).}
  \label{areacorrel}
\end{figure}

Once the events having their origin in the PMTs have been studied
and characterized (see section \ref{withoutsamples}), the
measurements taken with the samples of natural quartz, synthetic
quartz and methacrylate will be analyzed in this section. Figure
\ref{areacorrel} shows the distributions of the Asy parameter for
the three materials and two measurements conditions; the same
distributions in the measurements without sample (already shown in
figure~\ref{PMTsasymmetry}) are depicted also for comparison.

\begin{figure}
 \begin{center}
   \includegraphics[width=8cm]{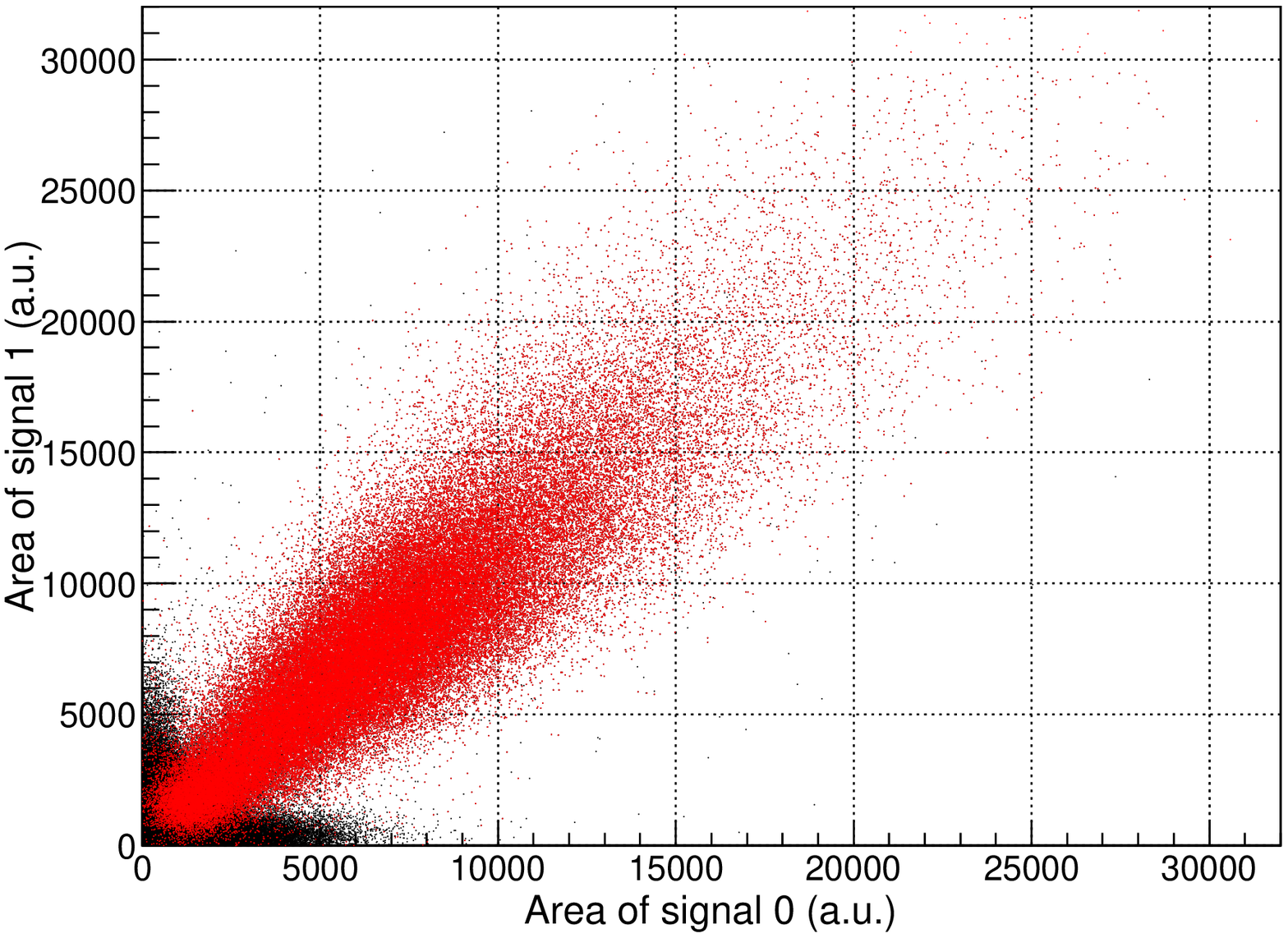}
 \end{center}
 \caption{Correlation between the areas of each PMT signal for the natural quartz sample in the background measurement. Results before (in black) and after (in red) rejecting PMT-like events are shown.}% (left) and under exposure to a $^{232}$Th source (right).}
  \label{areacorrelfil}
\end{figure}

%background: ritmo, correlaciones.
Following table~\ref{measurements}, the trigger rate in background
measurements is significantly higher than in the measurement without
samples only for natural quartz, with an increase of about 30\%. In
the case of methacrylate, the rate is even reduced, reaching the
same level than in the measurement with the black cardboard; it
seems that the 10-cm thick methacrylate is somehow absorbing the
light produced by PMTs (as shown in figure \ref{areacorrel},
symmetric events are particularly suppressed respect to the
background measurement without sample). From the Asy distributions
in the background measurements presented in figure~\ref{areacorrel},
it is evident for natural quartz that a population of events having
correlated PMT signals is present; the correlated signals have in
addition large areas, as it can be seen in
figure~\ref{areacorrelfil}. For methacrylate and synthetic quartz
there is not such an evidence of large area correlated events.

\begin{figure}
 \begin{center}
   \includegraphics[width=7cm]{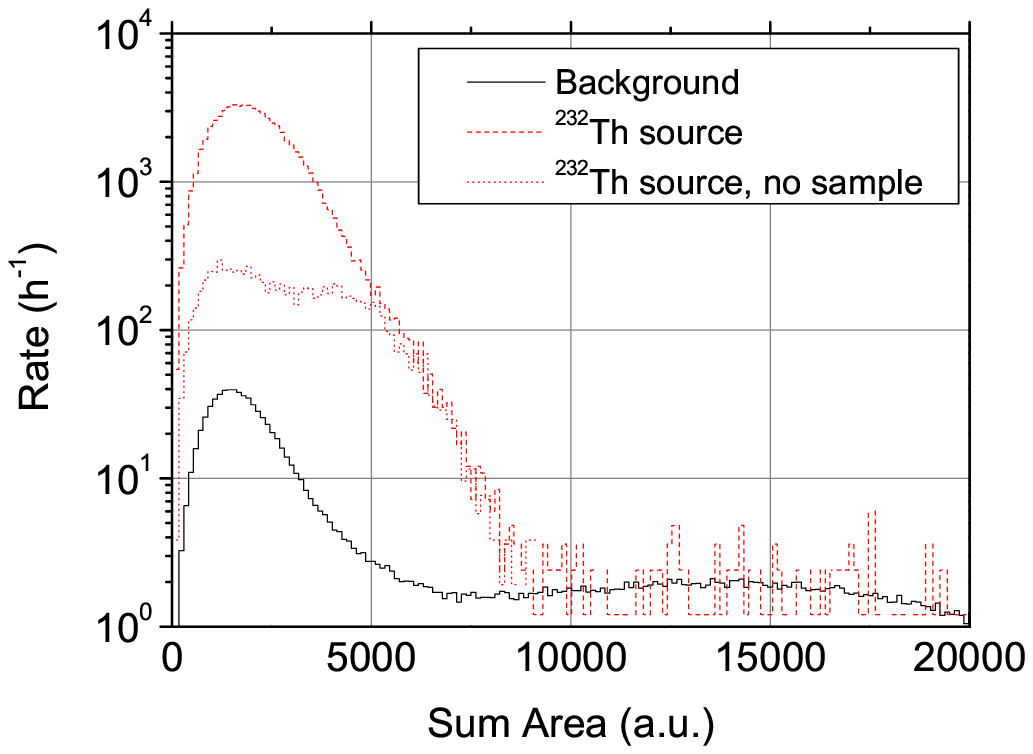}
      \includegraphics[width=7cm]{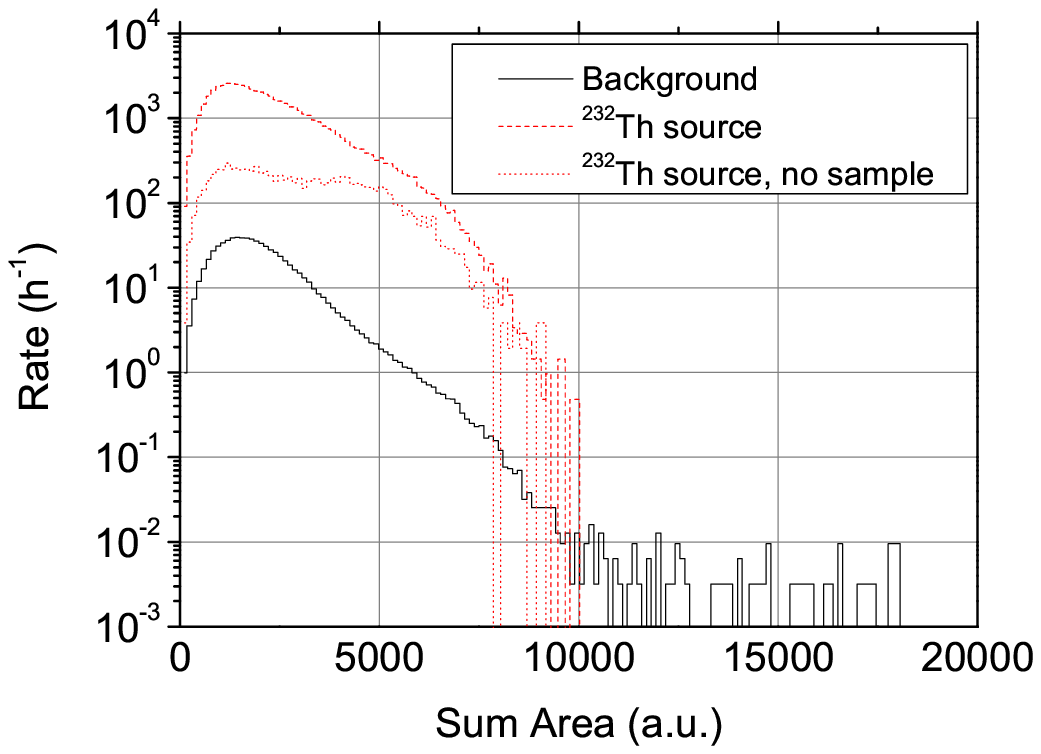}
         \includegraphics[width=7cm]{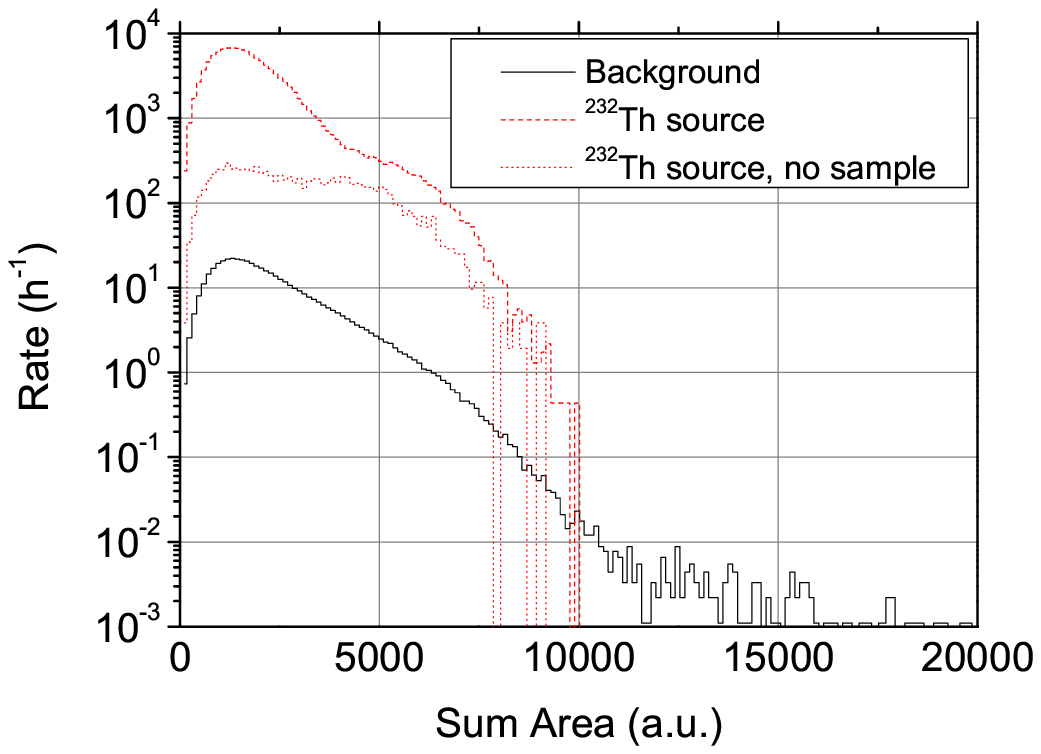}
 \end{center}
 \caption{Distributions (per hour) of the sum area of the PMT signals in the measurements with natural quartz, synthetic quartz and methacrylate (from top to bottom). Results for background (black solid lines) and under exposure to a $^{232}$Th source (red dashed lines) are compared. The distribution obtained without sample and with the $^{232}$Th is also depicted in all the plots (red dotted lines).}
  \label{areasTh}
\end{figure}

%fuente gamma: ritmo, correlaciones
For the three samples the external gamma source produces an
important increase in the trigger rates, comparing with the
background conditions, of almost two orders of magnitude in the
quartz samples and even higher for methacrylate (see
table~\ref{measurements}); following figure~\ref{areacorrel}, the
large amount of events generated under the gamma flux of the
$^{232}$Th source show a higher symmetry than those without the
samples, pointing to some kind of light generation mechanism
triggered by the gamma flux. Figure~\ref{areasTh} compares the
distributions of the sum area of the two PMT signals with the
$^{232}$Th source and in background conditions, for the three
samples; the distribution for the measurement without sample and
with the $^{232}$Th is also included in all the plots in order to
analyze the effect of the gamma flux on the sample. The rate
increases due to the gamma flux in general for all the area range,
but the bump that appears at high values (from 4000 to 8000 a.u.)
seems to be mainly related to interactions in the PMTs rather than
in the samples. However, for natural quartz the population of
correlated events with very large areas (above 10000 a.u.) presents
the same rate\footnote{The low statistics in the measurement with
the $^{232}$Th source is due to the short time of data taking, in
comparison with the background run.}.

\begin{figure}
 \begin{center}
   \includegraphics[width=13cm]{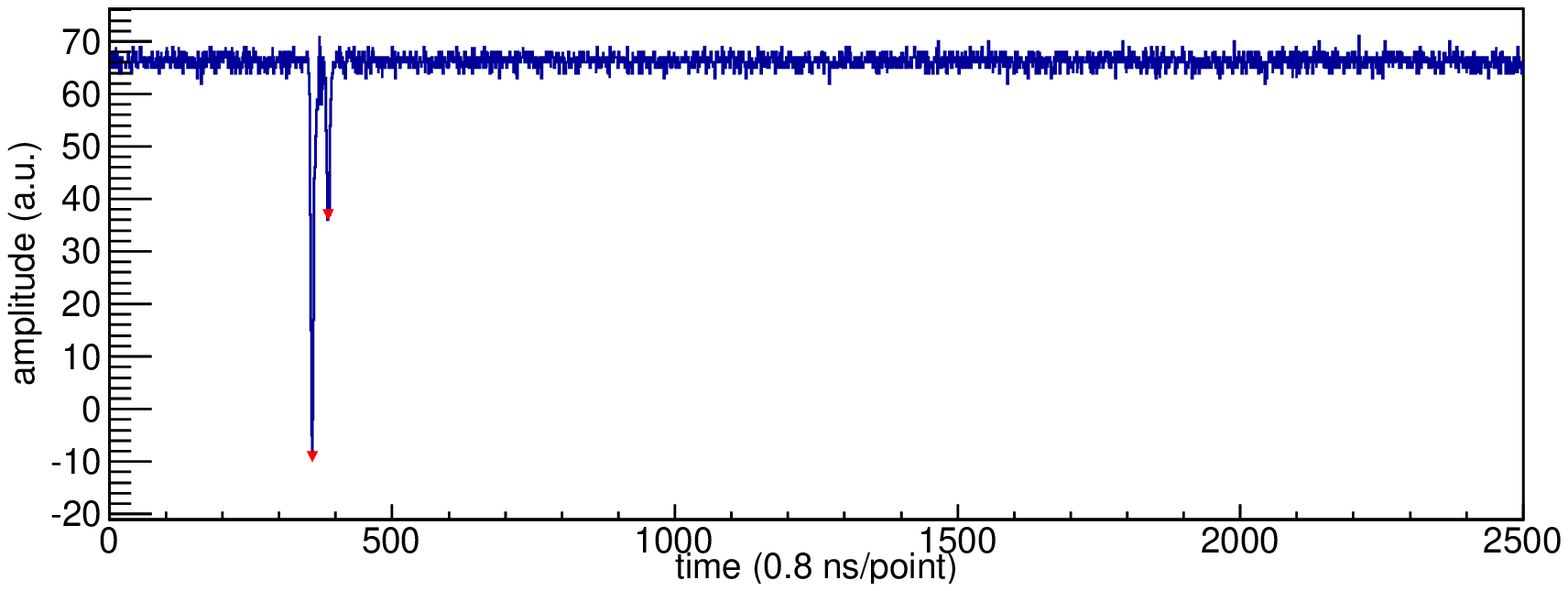}
    \includegraphics[width=13cm]{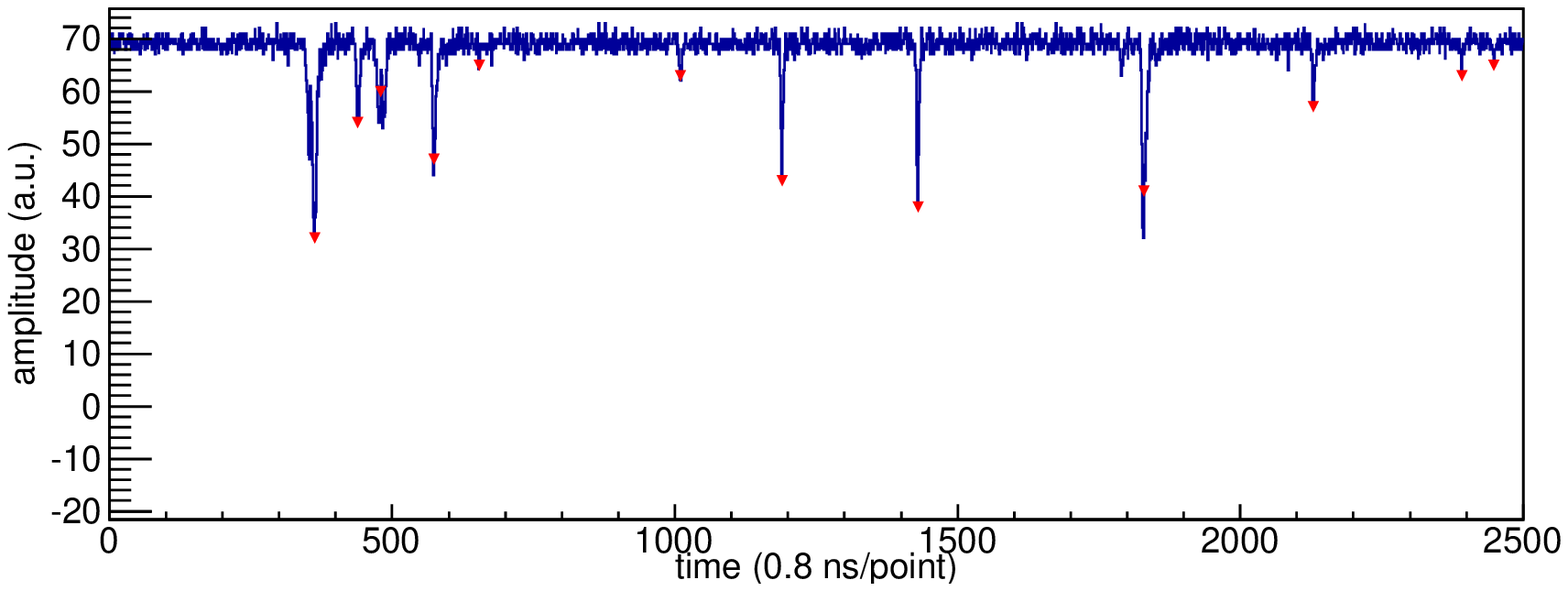}
    \includegraphics[width=13cm]{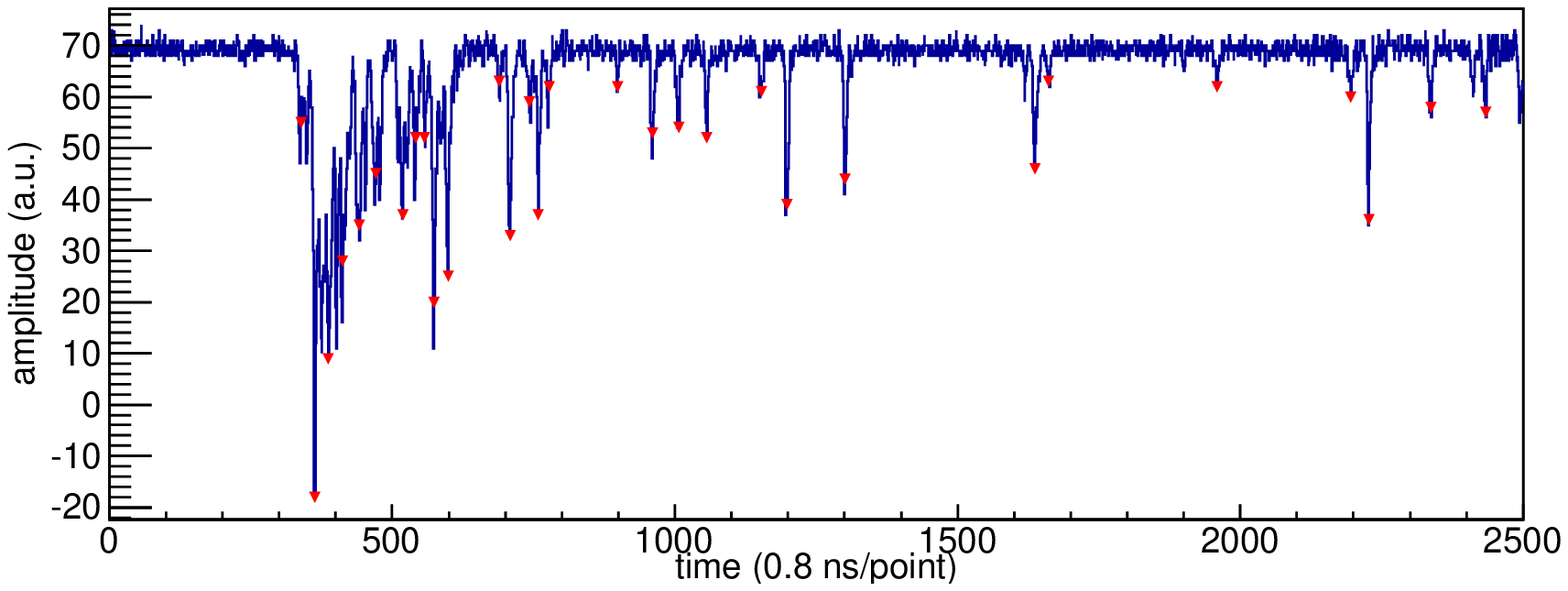}
 \end{center}
 \caption{One of the PMT signals for a PMT-like event obtained in the measurement with  $^{232}$Th source and black
 cardboard (top) and for two events with low and high values of n from the background measurement with the natural quartz (middle and bottom). The peaks identified in the pulses are marked with triangles and counted to derive the n parameter.}
  \label{examplepulsen}
\end{figure}

% descripci\'on filtrado
In order to disentangle the events which could be attributed to
scintillation in the samples from those just produced by PMTs, a
selection process has been applied to the background and $^{232}$Th
measurements with the three samples and without sample. Results at
different steps of this selection procedure are summarized in
table~\ref{filter}. As mentioned before, electronic noise is first
removed (step 1 in table~\ref{filter}) and quality cuts are then
applied to reject events showing baselines values out of range or
negative areas due to analysis artifacts (step 2 in
table~\ref{filter}). Finally, events compatible with a PMT origin
can be identified by counting the number n of peaks in the recorded
pulses and also considering the P1 values \cite{tesisclara} (step 3
in table~\ref{filter}). Figure \ref{examplepulsen} shows one of the
PMT signals of three example events chosen with very different n
values, one obtained without sample and attributed to PMT origin
(top) and the other two corresponding to the natural quartz
background measurement (middle and bottom). In particular, values
n$>$4 and P1$>$0.3 at both PMT signals, derived from the
distributions shown in figure~\ref{nP1}, have been required for
selecting non PMT-like events. The so-called PMT-like events include
not only PMT noise but also any fast light emission in the samples,
in the few ns range, which is indistinguishable from events having
the origin directly in the PMT. Hence, it has to be stressed that
only relatively slow scintillation events survive in the selection
process applied and can be singled out by this method.

\begin{table}
\begin{center}
\caption{Counting rates (h$^{-1}$) at different steps of the
selection process applied in the background measurements and with
the $^{232}$Th source for the three samples and without sample: 1)
after removing electronic noise, 2) after applying quality cuts, 3)
after filtering PMT-like events. The rate of identified PMT-like
events corresponds to events in the step 2 minus those in step 3.
Only statistical uncertainties are reported.} \centering
\begin{tabular}{|l|c|c|c|c|}
\hline
  & Without sample &   Natural quartz&            Synthetic quartz&            Methacrylate    \\ \hline
Background & & & & \\ \hline
%0  &921.4$\pm$3.4 & 1218.0$\pm$2.0  & 940.4$\pm$1.7 &621.38$\pm$0.83   \\
1 & 920.9$\pm$3.4 & 1217.2$\pm$2.0 & 938.4$\pm$1.7 & 615.5$\pm$0.8\\
2 &632.9$\pm$2.8&   979.5$\pm$1.8&   744.1$\pm$1.5  &468.1$\pm$0.7 \\
3 & 0.05$\pm$0.03  &302.6$\pm$1.0 & 0.05$\pm$0.01 & 0.08$\pm$0.01 \\
PMT-like events & 632.9$\pm$2.8 &  676.9$\pm$2.0 &  744.1$\pm$1.5 &
468.1$\pm$0.7\\ \hline

$^{232}$Th source & & & & \\ \hline
% 0 & (12.79$\pm$0.16)$\times10^{3}$ & (79.69$\pm$0.31)$\times10^{3}$  &  (70.74$\pm$0.19)$\times10^{3}$ & (149.78$\pm$0.26)$\times10^{3}$\\
1 & (12.70$\pm$0.16)$\times10^{3}$ & (79.63$\pm$0.31)$\times10^{3}$
& (70.74$\pm$0.19)$\times10^{3}$ & (149.77$\pm$0.26)$\times10^{3}$\\
2 & (8.98$\pm$0.13)$\times10^{3}$ & (69.04$\pm$0.29)$\times10^{3}$
& (57.55$\pm$0.17)$\times10^{3}$ & (117.53$\pm$0.23)$\times10^{3}$ \\
3  & $<$5.8 (95\%C.L.) & 237$\pm$17&  0.96$\pm$0.68& 0.44$\pm$0.44\\
PMT-like events & (8.98$\pm$0.13)$\times10^{3}$ &
(68.80$\pm$0.29)$\times10^{3}$ &
(57.55$\pm$0.17)$\times10^{3}$ & (117.53$\pm$0.23)$\times10^{3}$\\
\hline

$^{232}$Th source & & & & \\
with black cardboard  & & & & \\ \hline
%0 &656$\pm$17  & & &  \\
1 &654$\pm$17  & & &  \\
2 &568$\pm$16  & & &  \\
3 &$<$1.4 (95\%C.L.)  & & &  \\
PMT-like events &568$\pm$16  & & &  \\\hline
\end{tabular}
 \label{filter}
\end{center}
\end{table}

\begin{figure}
 \begin{center}
 \includegraphics[width=8cm]{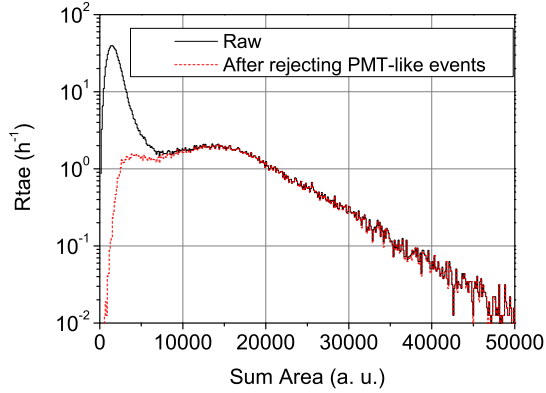}
 \end{center}
 \caption{Distribution (per hour) of the sum of the areas for
 the natural quartz sample in the background measurement. Results before (black solid line) and after (red dashed line) rejecting PMT-like events are shown.}% (left) and under exposure to a $^{232}$Th source (right).}
  \label{areafil}
\end{figure}

%filtrado en fondos
In the background measurements very few events remain after the
filtering of PMT-like events, except for natural quartz: in this
case, a population of events is singled out by the filtering,
showing a rate almost four orders of magnitude higher than the
corresponding to the other materials. This population can be
attributed to scintillation in natural quartz, that will be below
further analyzed. It is worth noting that according to
table~\ref{filter} the PMT-like events rate is of the same order in
the four cases, which might be considered a confirmation of their
common origin. In figure \ref{areacorrelfil} the effect of the
selection process on the correlation between the areas of each PMT
signal is presented for the background measurement with the natural
quartz sample; the surviving population of large, symmetric events
in natural quartz is clearly singled out.
%distribuciones de areas
Distributions of the event areas summing the contributions from the
two PMT signals, before and after applying the selection process,
are presented also in figure \ref{areafil} for the natural quartz in
the background measurement; the region of large area events is
unaffected.

%filtrado con fuente
Similar results are obtained when applying the described selection
process also to the measurements performed with the $^{232}$Th
source. For the natural quartz sample 197~events remain,
corresponding to a rate of (237$\pm$17)~h$^{-1}$. This rate is of
same order, even lower, than the rate of (302.6$\pm$1.0) h$^{-1}$
obtained in the background measurement, which indicates that the
gamma excitation does not influence the population of scintillation
events identified in natural quartz. This analysis should have
produced more compatibility between both estimates of scintillation
events; being this discrepancy an estimate of the possible
systematic effects, very difficult to evaluate, in the application
of the cuts (see table \ref{filter}).

%\begin{figure}
% \begin{center}
% \includegraphics[width=10cm]{areasrawfilter}
% \end{center}
% \caption{Distribution (per time unit) of the sum of the areas of the two PMT signals, before (in black) and after
%(in red) rejecting PMT-like events, in the background measurement
%for natural quartz.}
%  \label{areadist}
%\end{figure}

The presented results altogether confirm the existence for the
natural quartz of a population of events having large and highly
positively correlated areas in both PMT signals, with a number of
peaks per event pulse and a P1 value which are not compatible with
those of PMT-like events.

It has been already commented that for all the materials strong
increase in rate is observed under gamma irradiation. Events
contributing to this rate increase are indistinguishable from PMT
noise, according to our analysis. These events could be explained by
any prompt light generation mechanism, as for instance will be
expected for the Cherenkov emission in quartz, methacrylate or even
the PMT glass: corresponding energy threshold, timing properties and
light yield expected \cite{knoll} are roughly consistent with the
observations. Since an event-by-event selection of the populations
induced by the gamma flux has not been possible, in next section we
will focus in further understanding of the natural quartz
scintillation.

\subsection{Characterization of scintillation in natural quartz}

\begin{figure}
 \begin{center}
  \includegraphics[width=8cm]{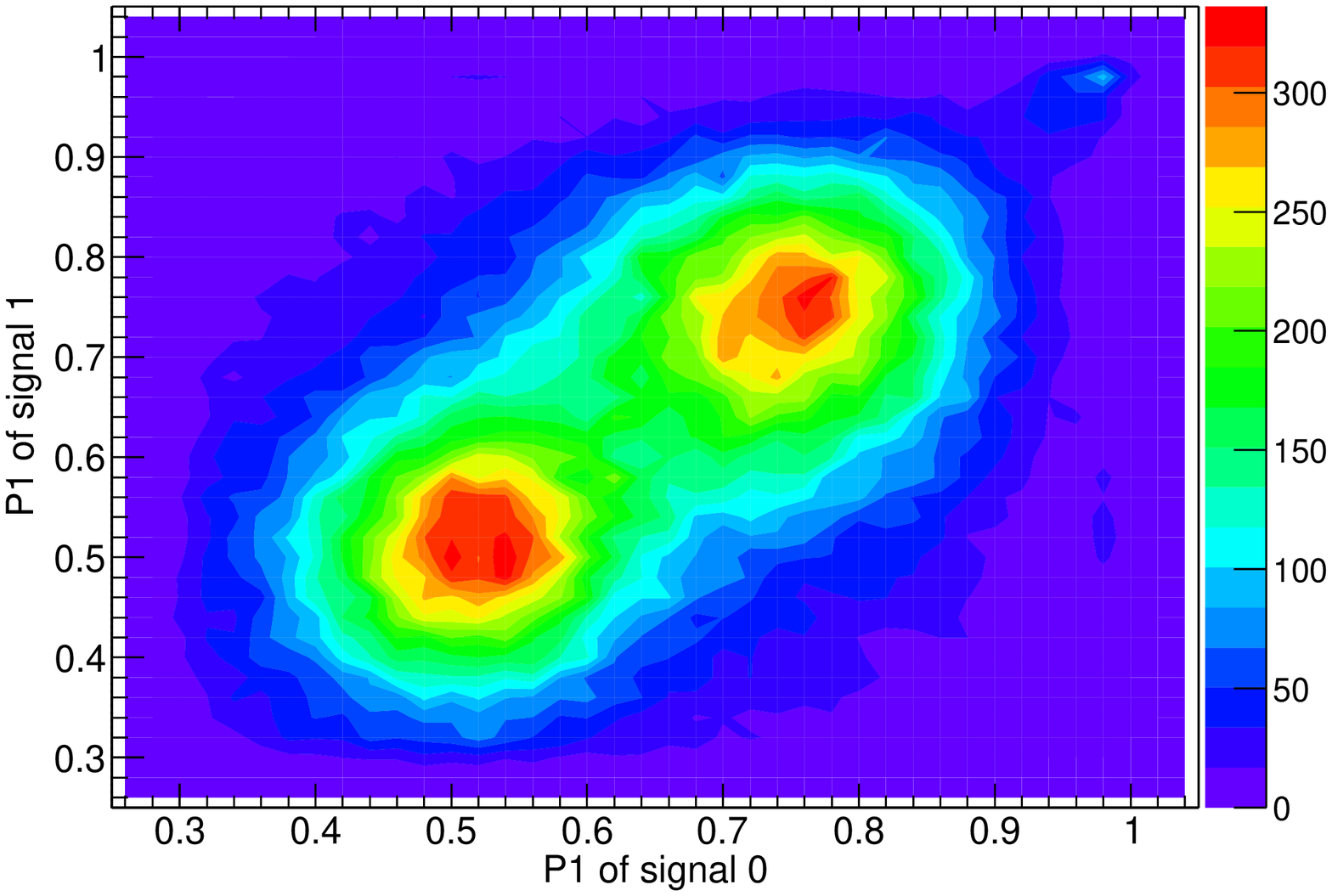}
    \includegraphics[width=8cm]{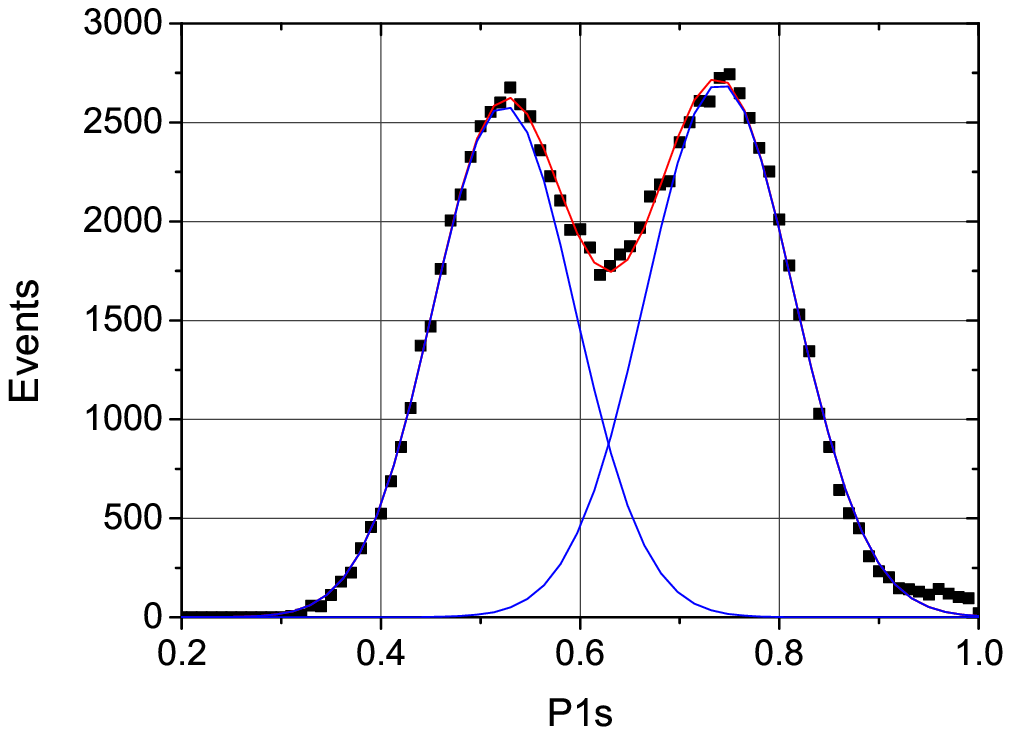}
 \end{center}
 \caption{Correlation between the P1 values of the two PMT signals (top) and distribution of P1s parameter (bottom), for the events attributed to scintillation in the background measurement with natural quartz, after filtering PMT-like events. Two different populations can be distinguished, referred as population I (II) the one with the faster (slower) events. The fit to two gaussians of the P1s distribution is depicted in solid lines and results shown in table~\ref{fmvalues}.}
  \label{fm}
\end{figure}

%separacion de poblaciones
The presence of two possible different populations of scintillation
events in natural quartz with large and positively correlated areas
is evidenced by analyzing the P1 parameter distribution for the two
PMT signals, shown in figure~\ref{fm}, top. We will refer in the
following to population I (II) as the one corresponding to faster
(slower) events with the lower (higher) P1 values.
%To separate the two populations, the line y=1.7-1.7*x in the shown plane has been considered.
A two-gaussian fit has been attempted for the P1s distribution,
shown in figure~\ref{fm}, bottom; the fit parameters are summarized
in table~\ref{fmvalues}. The areas of each gaussian give the number
of events assigned to each population, which are roughly half of the
total number of non PMT-like events.

%\begin{figure}
% \begin{center}
%  \includegraphics[width=10cm]{distributionP10s}
% \end{center}
% \caption{Distribution of P1s parameter for events shown in figure~\ref{fm}; gaussian fit results are depicted in solid lines.}
%  \label{p1s}
%\end{figure}

\begin{table}
\begin{center}
\caption{Results of the two-gaussian fit (plotted in figure
\ref{fm}, bottom) of the distribution of the P1s parameter for the
populations singled out in natural quartz.} \centering
\begin{tabular}{|l|c|c|}
\hline & Population I &  Population II  \\ \hline

$\mu$ & 0.5237$\pm$0.0008 & 0.7404$\pm$0.0008  \\
$\sigma$ & 0.0711$\pm$0.0008 &  0.0745$\pm$0.0008 \\
%Area & (460.7$\pm$4.8)$\times10^{2}$ & (503.6$\pm$4.9)$\times10^{2}$
Area rate (h$^{-1}$) & 144.4$\pm$1.5 & 157.8$\pm$1.5
\\ \hline

\end{tabular}
 \label{fmvalues}
\end{center}
\end{table}

\begin{figure}
 \begin{center}
 \includegraphics[width=6cm]{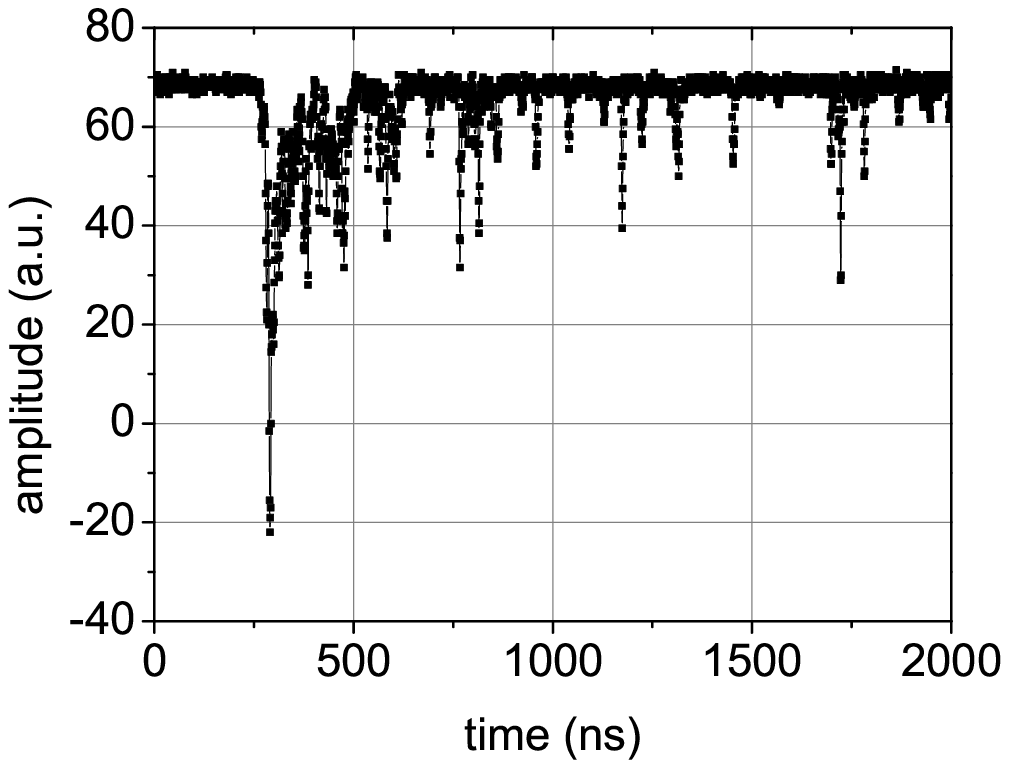}
  \includegraphics[width=6cm]{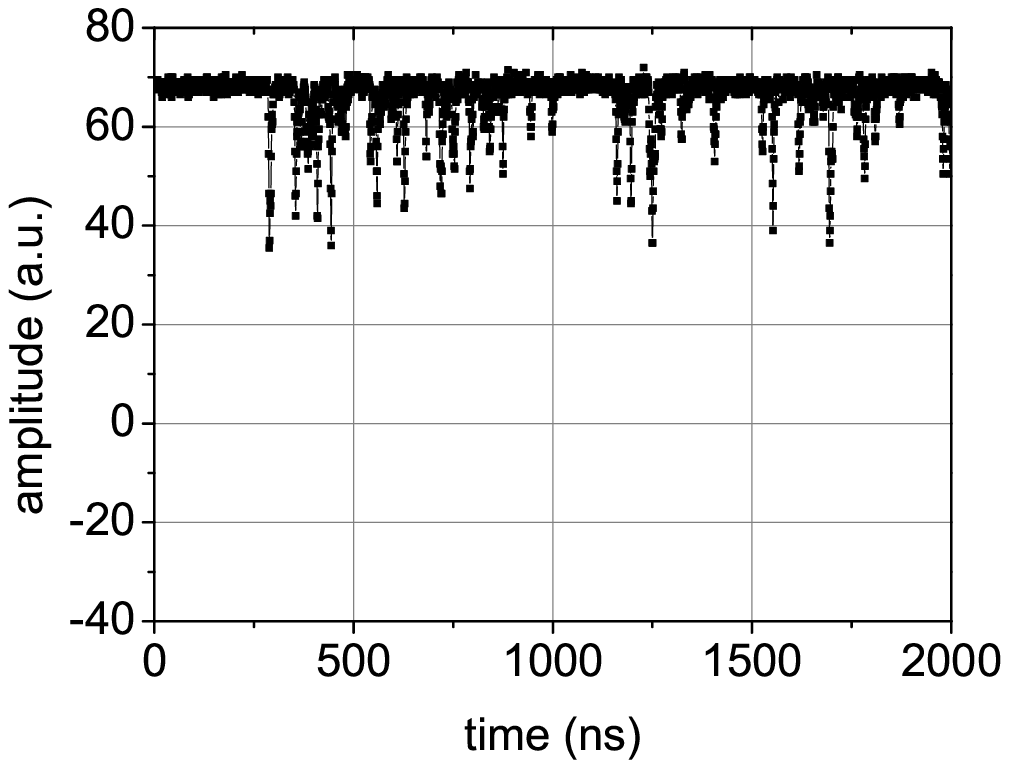}
  \includegraphics[width=6cm]{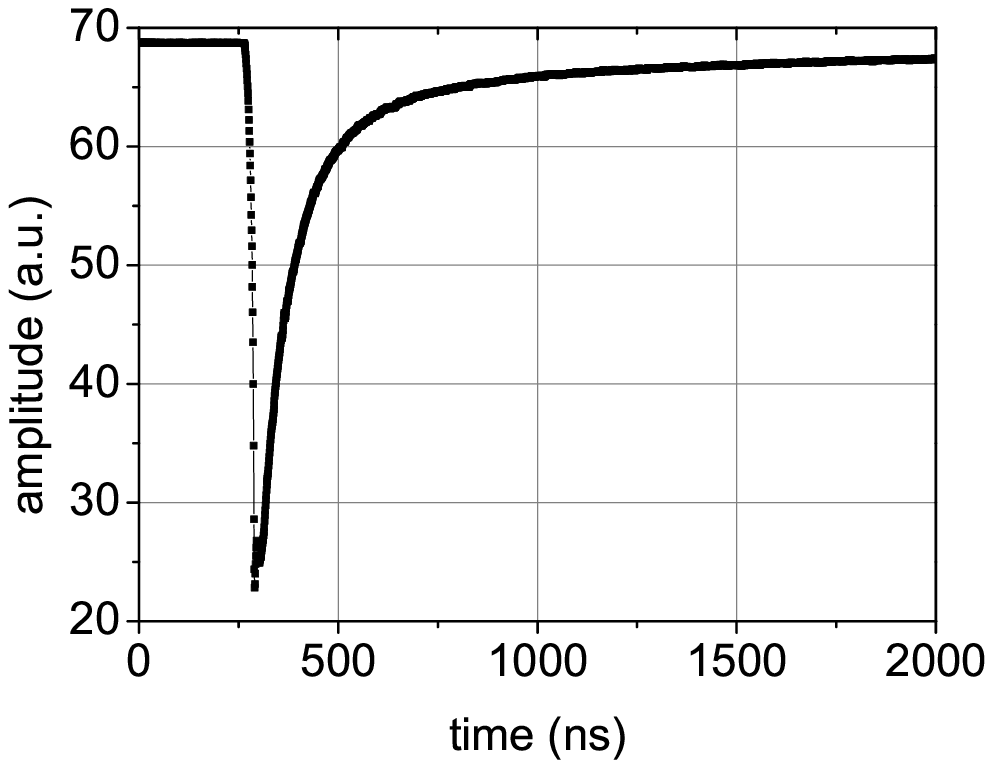}
  \includegraphics[width=6cm]{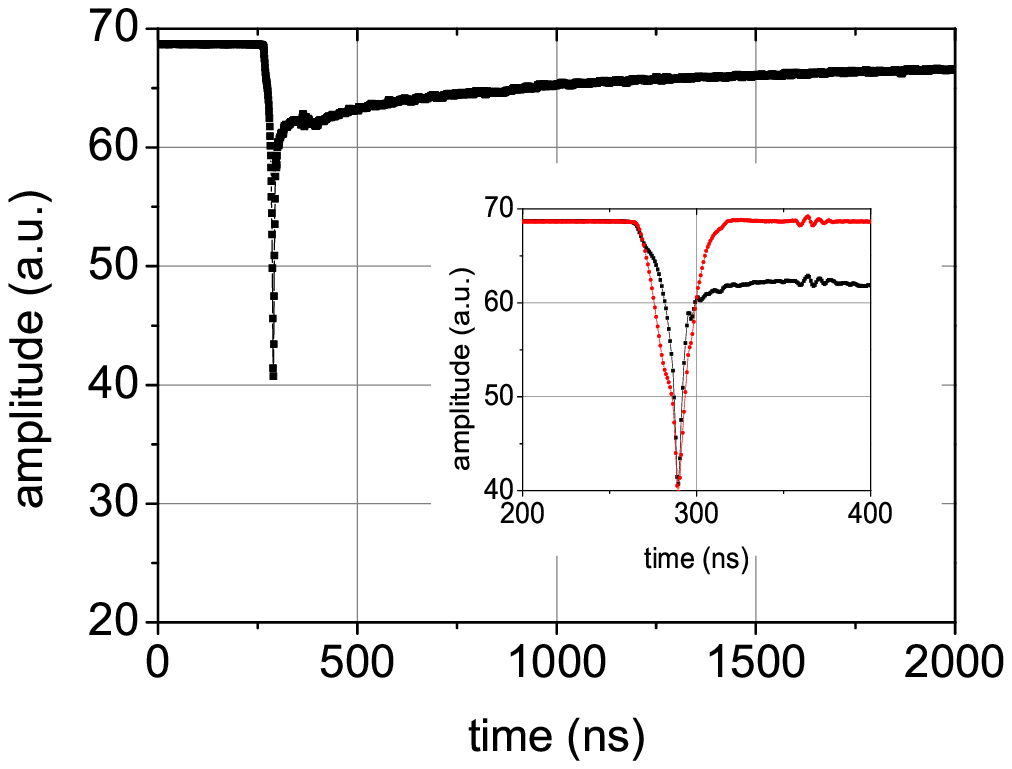}
 \end{center}
 \caption{Examples of individual pulses (top) and the mean pulse shapes (bottom) derived for the population I (left) and population II (right) of scintillation events identified in natural quartz. Average of the two PMT signals have been considered for the building of the mean pulse shapes. The inset in the plot of the mean pulse of population II zooms the beginning of the pulse and shows the comparison with the mean pulse obtained for single photoelectrons (selected with n$=$1) in the background measurement without sample (red line).}
  \label{pulses}
\end{figure}

Pulse shapes are quite different for both populations. To consider
the most representative events of each population, only events in
the 1$\sigma$ window around the mean value in the P1s distributions
of figure~\ref{fm}, bottom, have been taken into account to derive
mean pulses. Figure~\ref{pulses} presents these mean pulses as well
as a typical individual pulse for populations I and II; the two PMT
signals are well compatible and therefore have been averaged. This
average pulse has been fit to a sum of exponential decay terms in
order to deduce possible scintillation time constants:
\begin{equation}
V(t)=DC + A_{rise} \exp(-\frac{t-t_{0}}{\tau_{rise}}) - \sum A_{i}
\exp(-\frac{t-t_{0}}{\tau_{i}}) \label{pulso}
\end{equation}
being DC the baseline value, $t_{0}$ the start time of the pulse,
$A_{rise}$ and $\tau_{rise}$ the amplitude and time constant of the
rise component and $A_{i}$ and $\tau_{i}$ those of the i-th decay
term.

\begin{figure}
 \begin{center}
 \includegraphics[width=8cm]{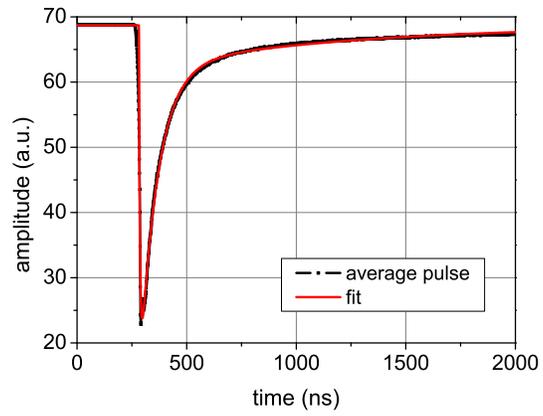}
   \includegraphics[width=8cm]{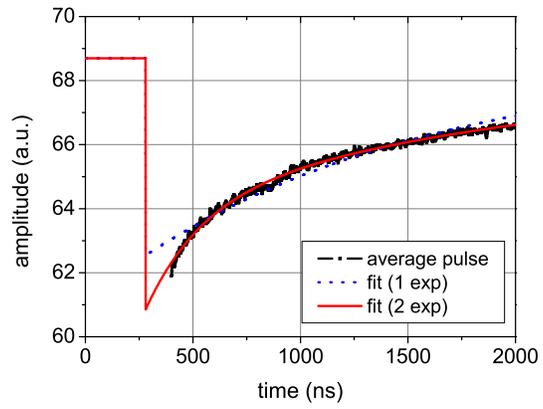}
 \end{center}
 \caption{Average pulses (in black) and the corresponding fits to eq.~(\ref{pulso}) for scintillation events populations I (top) and II (bottom) identified in natural quartz; fits with two decay time constants (in red) or only one (in blue) are shown. Fit parameters are summarized in table~\ref{fitres}.}
  \label{fitpop}
\end{figure}

%%ajuste pI
Figure~\ref{fitpop}, top presents the average pulse shape for
population I events (already shown in figure~\ref{pulses}) and the
corresponding best fit found, using eq.~(\ref{pulso}) with one rise
and two decay time constants. The results of the fit are summarized
in table~\ref{fitres}. The rise time constant is of the order of the
one expected for the PMT model used while the decay time constants
are in the typical range for inorganic scintillators; for example,
for oxides with Si time constants reported in \cite{lecoq} range
from some tens to some hundreds ns. The decay time constants found
are fully compatible with those derived when fitting only the decay
of the pulse to two exponential components.

%%ajuste pII
As shown in figure~\ref{pulses}, bottom right, the mean pulse for
population II presents a strange behavior in the first nanoseconds
range. A zoom of this region is shown in the inset of
figure~\ref{pulses} and the mean pulse for population II is compared
with the mean pulse corresponding to single photoelectrons (selected
by choosing n$=$1 in the data from the background measurement
without sample); the similarity between the two pulse shapes points
to an artifact produced at the averaging by the fixed trigger
position and strengthened by the low photon density registered in
population II events.
%due to the fact that only events with a common $t_{0}$ have been considered to sum them properly.
%Unlike for population I, a large dispersion in $t_{0}$ values deduced in the analysis for each pulse has been found in population II.
Therefore, only the decay part of the pulse, from 400~ns, has been
considered in the fit. The best fits found for one or two decay time
constants are shown in figure~\ref{fitpop}, bottom, and the results
presented in table~\ref{fitres}. The digitization window of 2~$\mu$s
is showing just the beginning of a much longer signal, then our data
cannot allow to derive the decay time constants precisely: the fit
to only one exponential decay does not explain properly our
observations (see figure~\ref{fitpop}, bottom), whereas the fit to
two exponential decays is affected by strong uncertainties in the
fit parameters (see table~\ref{fitres}).

\begin{table}
\begin{center}
\caption{Results of the fits to eq.~(\ref{pulso}) of the average
mean pulses shown in figure~\ref{fitpop} for the two populations
identified in natural quartz, considering different exponential
terms: one rise and two decay constants for population I but only
one or two decay constants for population II.} \centering
\begin{tabular}{|l|cc|cc|cc|}
\hline & Population I & & Population II & & Population II& \\ \hline
& $A$ (a.u.) & $\tau$ (ns) & $A$ (a.u.) & $\tau$ ($\mu$s) & $A$
(a.u.) & $\tau$ ($\mu$s) \\\hline
Rise &  48$\pm$49& 5.0$\pm$0.9 & - & - & - & - \\

Decay 1& 49.1$\pm$3.6& 81.1$\pm$6.6 & 3.4$\pm$2.9 & 0.27$\pm$0.66 &   6.16$\pm$0.59& 1.40$\pm$0.26 \\

Decay 2&6.6$\pm$1.2& 932$\pm$205&  4.4$\pm$4.7& 2.3$\pm$3.7 & - & -
\\ \hline
$t_{0}$ (ns)& 283.0 $\pm$  5.1 & & 280.0 (fixed) & & 280.0 (fixed) & \\
DC (a.u.) & 68.7 (fixed) & & 68.7 (fixed) & & 68.7 (fixed) & \\
\hline

\end{tabular}
 \label{fitres}
\end{center}
\end{table}

\begin{figure}
 \begin{center}
\includegraphics[width=8cm]{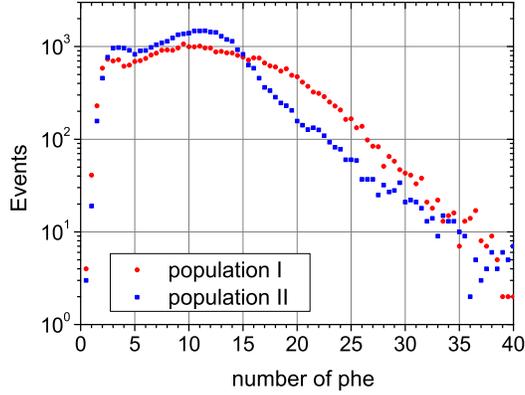}
 \end{center}
 \caption{Distributions of the number of photoelectrons yielded for the populations I (red circles) and II (blue squares) identified in natural quartz,
 estimated as the sum area of the PMT signals relative to the mean sum area of pulses with n$=$1 in the background measurement without sample.} \label{areaP1}
\end{figure}

%%yields
A rough estimation of the scintillation yield in the natural quartz
can be attempted comparing the areas of the selected events for the
two populations with those obtained without sample. Figure
\ref{areaP1} shows the distributions of the sum area of the PMT
signals relative to the mean sum area of pulses with n$=$1 in the
background measurement without sample, which we assume is an
estimate of the single photoelectron area. The deduced light yield
is about 12 (11) photoelectrons for populations I (II), although in
the latter only a fraction of the light emitted is registered
because of the digitization window chosen.

%Table~\ref{n} shows the number n of photoelectrons per event
%considering altogether the two PMT signals, estimated as the center
%of a gaussian distribution.
%\begin{table}
%\begin{center}
%\caption{Number n of photoelectrons per event for the two identified
%populations of possible scintillation in natural quartz and for the
%measurement without sample.} \centering
%\begin{tabular}{ccc}
%\hline Population I & Population II & without sample  \\ \hline
% 50.52$\pm$0.09 &  61.75$\pm$0.15 & 3.55 $\pm$0.01
%\\ \hline
%\end{tabular}
% \label{n}
%\end{center}
%\end{table}

\subsection{Possible interpretation} \label{int}
Once the effect observed in natural quartz has been characterized, a
possible explanation could be searched for. The first question to be
addressed is why luminescent centers are present in the natural
material while not in the synthetic one, and the second one, which
particles or emissions are responsible for the excitation of these
centers and the subsequent production of scintillation light.
%For the samples considered and according to the provider specifications,
%natural quartz darkens under UVA, X-ray and $\gamma$ irradiation,
%while synthetic quartz has no color change.

With respect to the first of the above proposed questions, in an
inorganic solid, luminescent centers are typically due to
substitutional impurities, excess atoms or ions and structure
defects \cite{librobirks,lecoq}. For instance, metallic impurities
are related to broken Si-O-Si bonds in quartz \cite{buchal}.
According to the chemical purity specifications given by the
provider, the highest metallic impurities in the quartz samples
measured are those of Al, with a value of 20~ppm for natural quartz
and 0.1~ppm for synthetic quartz. On the other hand, natural quartz
must have been more exposed to cosmic rays and environmental
radioactivity than synthetic quartz, and radiation exposure can
create defects in crystals (see for instance \cite{lecoq}).
%storing of energy from radiation is the base of the thermoluminescence and optical dating as mentioned before.

%In particular, formation of centers in natural quartz can be due to
%alpha or alpha-recoil particles from the radioactive elements of the
%U and Th series present in the material, as well as due to plastic
%deformation \cite{fukuchi}.
% In \cite{esr}, Al centers in SiO$_{2}$
%are reported to have optical absorption at 427 nm, a wavelength at
%the peak of the spectral response of the PMTs used in this
%experiment. The existence of a slow component in the scintillation,
%as suggested by the identified population II of events, could be
%attributed to the presence of oxygen ions in the material. Ions
%O$_{2}^{-}$ give typically green emissions with a decay time of
%several $\mu$s and can be produced by reactions under irradiation
%and/or temperature treatment \cite{lecoq}.

Concerning the second question, although in principle, different
particles could trigger the scintillation observed, we can discard
gamma radiation and cosmic muons and only internal alpha
contamination fulfills all the requirements:
\begin{itemize}
\item Environmental gamma radiation can be disregarded as the
rate of events in the two identified populations has not increased
when measuring with a $^{232}$Th source, producing photons of
different energies up to 2614.5~keV inside the lead shielding.
\item Cosmic muons cannot be responsible since the reduced flux
arriving at LSC cannot account for the rate of events measured for
the natural quartz sample by more than three orders of magnitude.
\item $\alpha$ particles generated by internal $^{232}$Th and $^{238}$U
contamination in the sample, with energies from about 4 to 8~MeV,
would be good candidates, specially taking into account results in
\cite{birks}. Assuming secular equilibrium in the chains and
considering that each alpha particle emitted gives a scintillation
event, an activity of 55~mBq/kg of $^{232}$Th or 41~mBq/kg of
$^{238}$U would be necessary to justify the counting rate of
scintillation events of 0.08~Hz measured for the two populations
altogether. Comparing these values with the activities reported in
table \ref{samples},
%for the quartz samples used, the upper limits for
%synthetic quartz are consistent with the fact of not observing
%scintillation effect like the reported for natural quartz in this material;
the measured value of $^{238}$U for natural quartz is larger than
necessary, but it must be noted that faint scintillation events
could have been lost in the PMT-like events rejection process and
also it may be possible that only alpha particles with energy above
a threshold produce a detectable effect. For synthetic quartz and
methacrylate only upper limits to the activity of $^{232}$Th and
$^{238}$U chains have been set, so the effect, even if not observed,
cannot be excluded.
%%%%%%%%% completar con medidas; si se confirman alfas, yield de 2-3 phe/MeV
\end{itemize}

\section{Conclusions} \label{conclusions}

Possible scintillation in natural and synthetic quartz and
methacrylate has been investigated at LSC, in a specially designed
scintillation test bench using two PMTs in coincidence. Measurements
have been carried out in low background environment as well as
exposed to an intense gamma flux generated by a $^{232}$Th source of
about 5 kBq.

The important increase in counting rates and the more correlated
signals registered by the two PMTs when exposing the materials to
the gamma flux points to the production, in all the investigated
samples, of a very fast light emission, which cannot be
distinguished from the PMT noise in our study. Hence, this kind of
events could be easily rejected when operating a detector with light
guides or coupling windows made of these materials. This effect is
probably due to the Cherenkov light emission by relativistic
particles, like electrons, produced in quartz, methacrylate and PMT
glass.

A different scintillation effect, observed in background conditions,
has been evidenced in natural quartz and partially characterized.
These events show large, highly correlated signals in both PMTs. Two
distinct populations have indeed been identified; one of them has
been fit to two exponential decays, obtaining values of the decay
time constants of the order of those of inorganic materials
((81$\pm$7)~ns and (0.9$\pm$0.2)~$\mu$s, the latter with smaller
amplitude) while the other population contains a slower component
with a decay time constant of $\sim$1-2~$\mu$s. This observed
scintillation in natural quartz optical windows could produce events
difficult to disentangle from scintillation at low energies, for
instance, in NaI(Tl) detectors.
%%%% conclusion sobre yield?

Although the origin of this scintillation has not been investigated
(and indeed two mechanisms with different properties seem to be
necessary to explain the two observed populations), the appearance
of luminescence centers in natural quartz could be attributed to
either impurities or to structural defects. Disregarded as origin of
the scintillation gamma radiation and cosmic muons, the hypothesis
of $\alpha$ particles from internal contaminations in the natural
chains seems the most plausible.

\section{Acknowledgements}
This work has been supported by the Spanish Ministerio de Econom\'ia
y Competitividad and the European Regional Development Fund
(MINECO-FEDER) (grants FPA2008-03228, FPA2011-23749), the
Consolider-Ingenio 2010 Programme under grants MULTIDARK CSD2009-
00064 and CPAN CSD2007-00042, and the Gobierno de Arag\'on (Group in
Nuclear and Astroparticle Physics, ARAID Foundation and C. Cuesta
predoctoral grant). C. Ginestra and P. Villar have been supported by
the MINECO Subprograma de Formaci\'on de Personal Investigador. We
also acknowledge LSC and GIFNA staff for their support.

\bibliographystyle{elsarticle-num}

\end{document}